\documentclass[12pt]{article}  
\usepackage{amsmath,amsfonts,amssymb}
\usepackage{graphicx}
\usepackage{setspace}
\usepackage{tocloft}
\usepackage{comment}
\usepackage{multirow}
\usepackage{colortbl}
\usepackage{url}
\title{SegTHOR: Segmentation of Thoracic Organs at Risk in CT images}

\author{Zo\'e Lambert$^a$, Caroline Petitjean$^b$, Bernard Dubray$^{b,c}$, Su Ruan$^b$\\ $^a$LMI, INSA Rouen, France \\$^b$Normandie Univ, INSA Rouen, UNIROUEN, \\ UNIHAVRE, LITIS, France \\ $^c$Centre Henri Becquerel, rue d'Amiens, 76000 Rouen, France}
\date{}

\cftpagenumbersoff{figure}
\cftpagenumbersoff{table}
\begin{document}
\maketitle

\begin{abstract}

In the era of open science, public datasets, along with common experimental protocol, help in the process of designing and validating data science algorithms; they also contribute to ease reproductibility and fair comparison between methods. Many datasets for image segmentation are available, each presenting its own challenges; however just a very few exist for radiotherapy planning. This paper is the presentation of a new dataset dedicated to the segmentation of organs at risk (OARs) in the thorax, i.e. the organs surrounding the tumour that must be preserved from irradiations during radiotherapy. This dataset is called SegTHOR (Segmentation of THoracic Organs at Risk). In this dataset, the OARs are the heart, the trachea, the aorta and the esophagus, which have varying spatial and appearance characteristics. The dataset includes 60 3D CT scans, divided into a training set of 40 and a test set of 20 patients, where the OARs have been contoured manually by an experienced radiotherapist. Along with the dataset, we present some baseline results, obtained using both the original, state-of-the-art architecture U-Net and a simplified version. We investigate different configurations of this baseline architecture that will serve as comparison for future studies on the SegTHOR dataset. Preliminary results show that
room for improvement is left, especially for smallest organs.
\end{abstract}




\section{Introduction}

Radiation therapy is one of the standard treatments for lung and esophageal cancer. It consists of irradiating the tumor with ionizing beams to prevent the proliferation of cancer cells. The goal is to destroy the target tumor while preserving healthy tissues and surrounding organs, called Organs at Risk (OARs), from radiation. Thus, delimiting the target tumor and OAR on computed tomography (CT) images is the first step in treatment planning. This segmentation task is mainly performed manually by an expert who relies on his experience and some medical guidelines.
In addition, manual segmentation is time-consuming and tedious. For these reasons, an automatic approach may be essential to improve and simplify the segmentation of OARs, and thus reduce the harmful effects of radiation therapy.

\begin{figure}[t]
    \centering
    \includegraphics[width=0.9\linewidth]{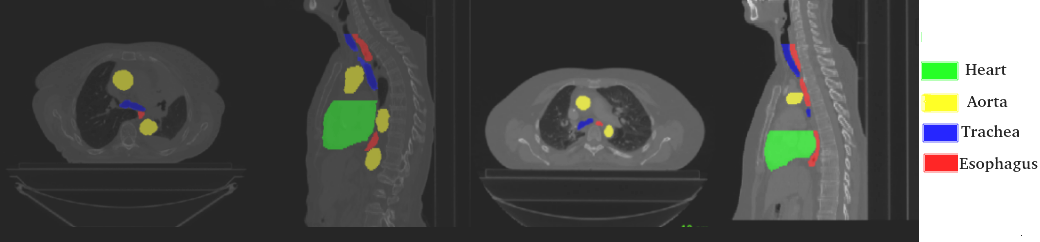}
    \caption{Example of CT image of 2 patients, with the axial (left) and sagittal (right) views, with an overlay of the manual segmentation of the 4 OAR. Figure best viewed in color.}
    \label{figct}
\end{figure}

In the spirit of making the segmentation of organs at risk automatic and more widely,
we have recently setup a dataset with data acquired at the Henri Becquerel Center (CHB), a regional anti-cancer center in Rouen, France. This data set, called SegTHOR for Segmentation of THoracic Organs at Risk, contains 60 CT scans from patients with lung cancer or Hodgkin's lymphoma. In this dataset, we focus on thoracic organs, which are heart, aorta, esophagus and trachea (Fig. \ref{figct}). These organs have varying shapes and appearances. The esophagus is the most difficult organ to contour due to its shape and position, which vary greatly from one patient to another and is almost invisible.

To the best of our knowledge, not many datasets exist with the purpose of organ at risk segmentation. The challenge\footnote{\url{http://aapmchallenges.cloudapp.net/competitions/3}} proposed by the AAPM (American Association of Physicists in Medicine) has a similar goal: it aims to segment the esophagus, heart, spinal cord, left and right lung in CT images. 30 patients are available as training set, while the test set includes the scans of 12 patients. The organs are different from the SegTHOR dataset: their dataset does not include trachea and aorta. Very recently, the StructSeg 2019\footnote{\url{https://structseg2019.grand-challenge.org/}} challenge proposes two segmentation tasks of OARs. The purpose of the first is to segment 22 OARs in head and neck CT scans from nasopharynx cancer patients. The second one aims to segment 6 OARs in chest CT scans from lung cancer patients. OARs are the same as those of the AAPM challenge with the trachea in addition. For both databases, 50 CT scans compose the training data and 10 others constitute the test data.

The goal of this paper is to present the SegTHOR dataset, and to give some baseline results, using the state-of-the-art segmentation networks. Note that this dataset has been the subject of a challenge  that we organized between January and April 2019\footnote{Due to a major crash of Codalab servers in July 2019, all results collected for the challenge have disappeared and are not accessible anymore. A new submission system has been setup and is accessible at \url{https://competitions.codalab.org/competitions/21145}.},  held at the IEEE International Symposium on Biomedical Imaging (ISBI) in April 2019 in Venice, Italy.

The paper is organized as follows. The dataset is described in the next section. 
In Section \ref{framework}, a brief overview of medical image segmentation is introduced to identify the state-of-the-art architectures; the proposed 2D network used for the automatic segmentation of thoracic organs on SegTHOR dataset is then presented. Results are reported in Section \ref{experiments}.

\section{The SegTHOR Dataset}\label{dataset}

The database consists of 60 thoracic CT scans, acquired with or without intravenous contrast, of 60 patients diagnosed with lung cancer or Hodgkin's lymphoma. These patients were treated by curative-intensive radiotherapy, between February 2016 and June 2017, at the Henri Becquerel Center (CHB, regional anti-cancer center), Rouen, France. All scanner images are 512$\times$512$\times$(150 $\sim$ 284) voxels in size. Indeed, the number of slices changes according to the patients. The in-plane resolution varies between 0.90 mm and 1.37 mm per pixel and the $z$-resolution fluctuates between 2 mm and 3.7 mm per pixel. Finally, the most common resolution is 0.98$\times$0.98$\times$2.5 mm$^3$.

Each CT scan is associated with a manual segmentation performed by an experienced radiotherapist at the CHB, using a SomaVision platform, Varian Medical Systems, Inc, Palo Alto, USA. Manual segmentation takes approximately 30 minutes for each patient. The body and lung contours were segmented with the automatic tools available on the platform. The esophagus was manually delineated from the fourth cervical vertebra to the esophago-gastric junction. The heart was delineated as recommended by the Radiation Therapy Oncology Group. The trachea was contoured from the lower limit of the larynx to 2cm below the carena excluding the lobar bronchi. The aorta was delineated from its origin above the heart down to below the diaphragm pillars.

The segmentation of these 4 OAR raises the following challenges. First, the tissues that are surrounding the heart and aorta, and specially the esophagus, have similar gray levels to these organs; the lack of contrast forces the radiotherapist to use his anatomical knowledge resulting in a segmentation that does not rely on the CT scan only. Note that the trachea, on the contrary, is easily identifiable because it is filled with air and thus appears as black on the image. Also, another challenge is the three-dimensional relationships of these OAR: they are intricately interlocked as shown in Fig. \ref{figct}. At last, the 4 OAR have varying shapes and size: esophagus and trachea have tubular structure and are the smallest organs; the aorta has a cane shape and the heart, the largest organ has a blob shape.

We have split the data in a training set of 40 patients and a test set of 20 patients, which represents 7390 slices for training data and 3694 slices for test data to define the SegTHOR dataset. The dataset is available at \url{https://competitions.codalab.org/competitions/21145}, with online automated evaluation. The Dice metric and the Average Hausdorf distance are provided for each OAR of the test set patients.

\section{A segmentation framework based on U-Net}\label{framework}

\subsection{Related work in medical image segmentation}\label{related}

Due to lack of contrast between the organs and surrounding tissues, the segmentation problem of OAR requires to rely on external knowledge, such as pairs of CT image and their corresponding manual labeling. Making use of prior knowledge and labeled images has been long used in medical image segmentation, to guide the segmentation process in case of noise and occlusion, and to handle object variability. For example, an atlas-based method, in addition to other techniques, was used to segment 17 OARs throughout the body \cite{han2015segmentation}. The segmentation of thoracic organs at risk is obtained in \cite{schreibmann2014multiatlas} by combining multi-atlas deformable registration with a level set-based local search.
In recent times, traditional image segmentation methods have been outperformed by convolutional neural networks (CNN)-based ones. One of the first CNN architectures to allow automatic end-to-end semantic segmentation is the Fully Convolutional Network (FCN) \cite{long2015fully}. FCN has paved the way for encoder-decoder segmentation networks. Among its successors, one of the most well-known  architecture is DeepLab \cite{chen2017deeplab}, where a combination of dilated convolutions and feature pyramid pooling is introduced.
The U-Net architecture \cite{u-net2015} is also a popular segmentation framework, initially designed for medical applications \cite{litjens2017survey}. It has a symmetrical encoder-decoder structure: the image is downsampled throughout the encoding path, and upsampled using transposed convolution (also called deconvolution) to reach the initial resolution. Some variants in U-Net consist in changing the backbone model used for encoding, e.g. VGG, DenseNet, etc. Extensions to 3D have been proposed in the 3D-UNet model \cite{3Dunet2016} and the V-Net model \cite{milletari2016v}.
For example, in \cite{roth2017hierarchical}  a multi-class 3D FCN is trained on CT scans to segment seven abdominal structures. In \cite{nikolov2018deep}, 21 OAR are segmented in the head and neck using a 3D-UNet architecture. The liver is segmented on CT images thanks to a 3D deeply supervised network in \cite{DBLP:journals/corr/Dou0JYQH16}, or to a hybrid densely connected UNet architecture in \cite{DBLP:journals/corr/abs-1709-07330}. In \cite{trullo2019multiorgan}, a distance map that provides the localization of each organ and the spatial relationship between them is used to guide the segmentation task in a fully convolutional setting.

\subsection{A simplified segmentation framework}

The U-Net architecture being the state-of-the-art model for image segmentation, our first intention is to evaluate this architecture \cite{u-net2015} on each 2D images of the SegTHOR test dataset. Given OAR contours high inter- and intra-patient variability, it is deemed to be subject to overfitting.  Our strategy has consisted in adapting U-Net to our problem by some simple steps.
The first step to tackle overfitting is to add dropout \cite{srivastava14a} regularization to the network, a common element in modern CNN. Dropout consists in randomly ignoring each neuron in the network with a probability \textit{p}, and therefore their connections, during each step of the training. This prevents the neurons from adapting too much to each other. A second way to reduce overfitting is to limit the number of network layers and feature maps, to reduce the number of trainable parameters. The result is a simplified architecture with one less hidden layer and only up to 256 feature maps calculated. Finally, we have chosen to replace the transposed convolution (also called deconvolution) by a bilinear interpolation for the upsampling operation, in the expansion phase. The first one requires learning the weights of the filters, while the second one uses neighboring pixels to calculate the value of the new pixel through linear interpolations, which further reduces the number of parameters.

\begin{figure}[t]
    \centering
    \includegraphics[width=0.9\textwidth]{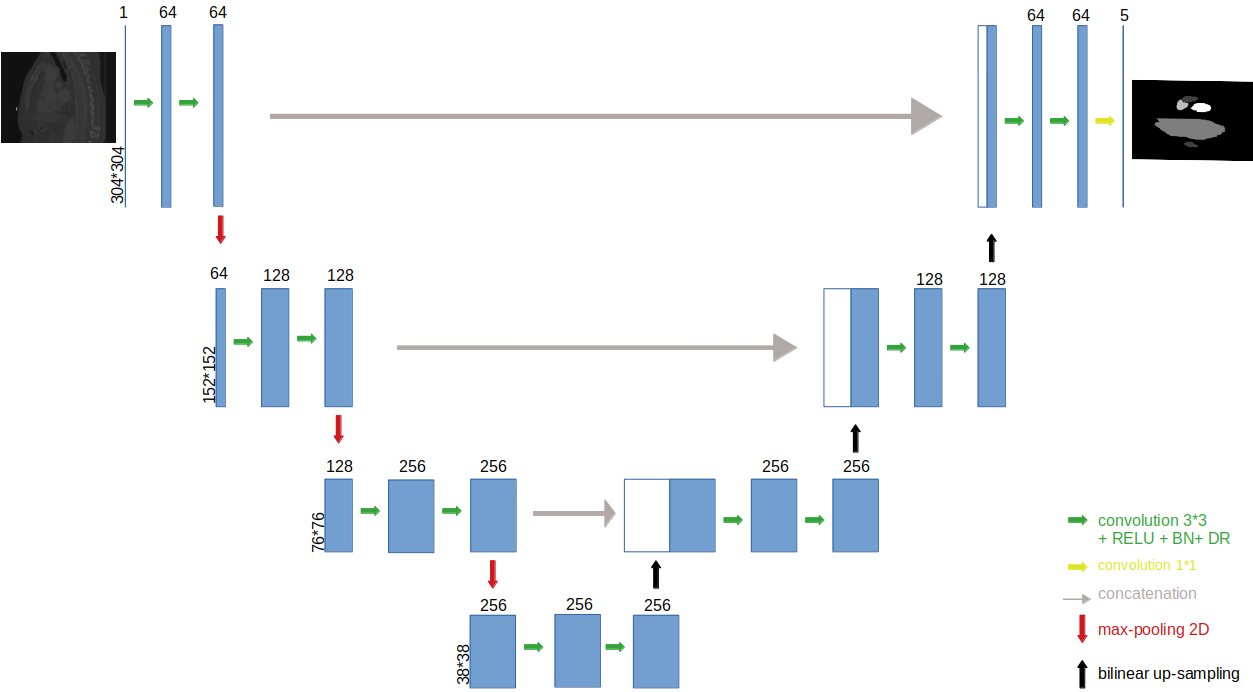}
    \caption{Architecture of the proposed simplified U-Net, denoted sU-Net. Each blue rectangle indicates a layer, where the number above is the number of feature maps. Corresponding image resolution is specified vertically.}
    \label{fig:reseau}
\end{figure}

As shown in Figure \ref{fig:reseau}, our simplified network has an encoder-decoder path composed of 7 convolutional blocks, some of which are connected by skip connections. Each convolutional block consists of two convolution operations with a 3$\times$3 kernel size. The ReLU (Rectified Linear Units) activation function and then a batch normalization are applied to the outputs of each convolution, along with a dropout. In the encoder part, the two convolution operations are followed by a max-pooling operation that reduces by half the spatial resolution of the input; while in the decoder part, the two convolution operations are preceded by a bilinear upsampling operation to double the spatial resolution and finally reach the initial resolution. Three skip connections are used to concatenate the characteristics of the first layers with those of the deeper ones to compensate for the loss of resolution. At the end of the network, there is a last convolution operation with a 1$\times$1 kernel size to obtain the feature maps associated to each segmentation classes, the background and the OARs. Finally, this architecture has 4.8 million trainable parameters compared to 7.2 million for the same architecture with the transposed convolution operation, while the original U-Net, based on a VGG backbone, has about 65 million trainable parameters.

\section{Experiments and result}\label{experiments}

\subsection{Pre-processing}

All images are normalized by subtracting the image mean and dividing by standard deviation. We increased the data to artificially triple the database size using data augmentation techniques. Each image is modified by a random affine transform on the one hand and a random deformation of a 2$\times$2$\times$2 control point grid and a B-Spline interpolation on the other hand \cite{milletari2016v,trullo2017}. For computational reasons, images are cropped from the center and are 304 $\times$ 304 pixels in size. In addition, only slices with at least one of the four organs are passed through the network during training. \\

\subsection{Implementation}

The four classes and background are highly unbalanced. Indeed, the background represents about 99\% of the voxels on average. The remaining percentage of voxels is divided into 70.7\% for the heart, 23\% for the aorta, 3.7\% for the esophagus and 2.6\% for the trachea. To overcome this problem, the multi-class Dice loss function, a generalization of the binary Dice loss function \cite{milletari2016v,carolesudre}, is used. It is optimized using the stochastic gradient descent algorithm with an initial learning rate of 1e$^{-3}$, over mini-batches of size 5. When learning no longer progresses, this learning rate is reduced by a factor of ten. Weight decay and momentum are set to 5e$^{-4}$ and 9e$^{-1}$, respectively. Finally, the weights in the network are initialized by Xavier's initialization. The deep network is implemented with PyTorch.

\subsection{Evaluation metrics}

To quantify our segmentation results, two metrics are used. First, the Dice score, which measures the overlap rate between manual and automatic segmentation. In complement to this metric, the Average Hausdorff distance (AHD) in mm is calculated as the maximum between average distances from manual to closest automatic contours and average distances from automatic to closest manual contours. These two scores are obtained for each of the four OARs.

\subsection{Results}

In the first experiment, we compare U-Net performance with the simplified sU-Net. We also assess the difference in segmentation accuracy, without and with dropout, with drop probability $p$ to 0.2. Next, we assess two different configurations in the decoder phase: (i) with a 2D transposed convolution operation (denoted conv2Dtranspose in the result table), and (ii)  with a bilinear upsampling operation, which are used to recover the initial resolution of the image. Whenever necessary, we assess the statistical significance of the results, by performing a Wilcoxon signed-ranked test on Dice values (as well as on AHD) between the two methods of interest, using a confidence interval of 95\%.\\

\begin{figure}
    \centering
    \includegraphics[width=0.9\textwidth]{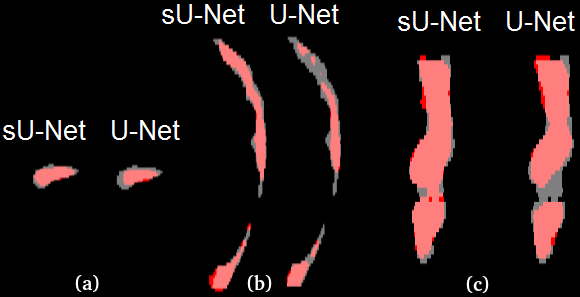}\\
    \caption{Comparison of segmentation results of the esophagus of sU-Net (left) and U-Net (right), for each view (a) axial, (b) sagittal and (c) coronal. Predicted areas are in red, while ground truth is in gray. 
    }
    \label{fig:comparison}
\end{figure}

\textbf{Comparison of sU-Net vs U-Net and influence of dropout.} Results are reported in Table \ref{results_dropout}. Comparing sU-Net to U-Net without dropout (columns (1) and (3)), it can be seen that results are similar. Now, if dropout is included in both networks, sU-Net shows enhanced performance compared to U-Net (columns (2) vs (4)), for all organs but the trachea. This is confirmed by the $p$-values of the Wilcoxon test, which are below the 0.05 threshold, for the esophagus, trachea, and aorta. Some qualitative results to illustrate the difference between the U-Net for the esophagus, are given in Figure \ref{fig:comparison}. The contribution of the dropout to the sU-Net framework can be assessed by comparing columns (3) and (4), where one can see that for 3 out of the 4 OAR, the dropout provides a substantial improvement, especially for the esophagus.\\



\begin{table}
\caption{Average segmentation results ($\pm$ standard deviation) for U-Net and our sU-Net framework with and without dropout (DR). Cell in yellow: sU-Net values that differ significantly from U-Net values, with DR (columns (4) vs (2)). In blue and bold: values with DR that significantly differ from without DR (columns (4) vs (3)).
}
\vspace{.5cm}
\label{results_dropout}
\begin{tabular}{|l|c||c|c||c|c|c|}
   \cline{3-6}
    \multicolumn{2}{l|}{} & \multicolumn{2}{c|}{U-Net} & \multicolumn{2}{c|}{Simplified U-Net (sU-Net)} \\
   \hline
    \multirow{2}{*}{OAR} & \multirow{2}{*}{Metrics} & without DR & with DR & without DR & with DR \\
        &  & \footnotesize{(1)} & \cellcolor{yellow}\footnotesize{(2)} & \textcolor{blue}{\textbf{\footnotesize{(3)}}}& \cellcolor{yellow}\textcolor{blue}{\textbf{\footnotesize{(4)}}} \\
    \hline
    \multirow{2}{*}{Esophagus} & \multirow{1}{*}{Dice} & \multirow{1}{*}{0.76 $\pm$  0.10} & \multirow{1}{*}{0.79 $\pm$ 0.08} & \multirow{1}{*}{0.75 $\pm$  0.11} & \cellcolor{yellow}\multirow{1}{*}{\textcolor{blue}{\textbf{0.82 $\pm$  0.05}}} \\
     \cline{2-2}
        & AHD & 1.74 $\pm$  2.77 & 0.94 $\pm$ 0.63  & 1.69 $\pm$  2.02 & \cellcolor{yellow}\multirow{1}{*}{\textcolor{blue}{\textbf{0.70 $\pm$ 0.39}}} \\
    \hline
    \multirow{2}{*}{Trachea} & \multirow{1}{*}{Dice} & \multirow{1}{*}{0.85 $\pm$  0.05} & \multirow{1}{*}{0.85 $\pm$ 0.04} & \multirow{1}{*}{\textcolor{blue}{\textbf{0.86 $\pm$  0.04}}} & \multirow{1}{*}{0.85 $\pm$ 0.04} \\
 \cline{2-2}
        & AHD & 1.32 $\pm$  1.20 & 1.30 $\pm$ 1.12 & 1.06 $\pm$  0.83 & 1.21 $\pm$  1.13 \\
        \hline
    \multirow{2}{*}{Aorta} & \multirow{1}{*}{Dice} & \multirow{1}{*}{0.92 $\pm$  0.05} & \multirow{1}{*}{0.91 $\pm$ 0.04} & \multirow{1}{*}{0.91 $\pm$  0.02} & \cellcolor{yellow}\multirow{1}{*}{0.91 $\pm$  0.03} \\
  \cline{2-2}
        & AHD & 0.50 $\pm$  0.64 & 0.77 $\pm$ 0.93 & 0.57 $\pm$ 0.65 & 0.58 $\pm$  0.67 \\
        \hline
    \multirow{2}{*}{Heart} & \multirow{1}{*}{Dice} & \multirow{1}{*}{0.93 $\pm$  0.03} & \multirow{1}{*}{0.93 $\pm$ 0.03} & \multirow{1}{*}{0.92 $\pm$  0.03} & \cellcolor{yellow}\multirow{1}{*}{\textcolor{blue}{\textbf{0.93 $\pm$  0.03}}}\\
    \cline{2-2}
        & AHD & 0.23 $\pm$  0.21 & 0.25 $\pm$ 0.28 &  0.31 $\pm$  0.22 & \textcolor{blue}{\textbf{0.27 $\pm$  0.20}} \\
        \hline
\end{tabular}
\end{table}

\textbf{Influence of upsampling method for the decoder.} Comparing the transposed convolution method and the bilinear interpolation method in Table \ref{results_upsample}, we find that the Dice and AHD values are not significantly different ($p \geq 0.05$) for all organs, but the aorta for which the $p$-value is 1.2e$^{-4}$ in favor of bilinear upsampling. Thus a bilinear upsampling operation is more than sufficient in this application. Moreover, choosing bilinear interpolation can help in reducing computation time.

\begin{table}
\caption{Average segmentation results ($\pm$ standard deviation) for U-Net and our sU-Net framework, both with DR. In blue and bold: values in sU-Net with bilinear upsampling that significantly differ ($p\geq 0.05$) from conv2Dtranspose (columns (3) vs (2)). }
\label{results_upsample}
\vspace{.5cm}
\begin{tabular}{|l|c||c||c|c|r|}
   \cline{3-5}
    \multicolumn{2}{l|}{} & \multicolumn{1}{c|}{U-Net} & \multicolumn{2}{c|}{Simplified U-Net (sU-Net)} \\
   \hline
    \multirow{2}{*}{OAR} & \multirow{2}{*}{Metrics}  & conv2Dtranspose &  conv2Dtranspose & bilinear upsampling \\
      &  & \footnotesize{(1)} &  \textcolor{blue}{\textbf{\footnotesize{(2)}}} & \textcolor{blue}{\textbf{\footnotesize{(3)}}} \\
    \hline
    \multirow{2}{*}{Esophagus} & Dice & 0.79 $\pm$ 0.08 & 0.82 $\pm$  0.05 & 0.81 $\pm$ 0.06\\
    \cline{2-2}
        & AHD & 0.94 $\pm$ 0.63  & 0.70 $\pm$  0.39 & 0.68 $\pm$ 0.35\\
    \hline

    \multirow{2}{*}{Trachea} & Dice  & 0.85 $\pm$ 0.04  & 0.85 $\pm$  0.04 & 0.86 $\pm$ 0.04  \\
    \cline{2-2}
        & AHD & 1.30 $\pm$ 1.12& 1.21 $\pm$  1.13 & 1.08 $\pm$ 0.85 \\
        \hline
    \multirow{2}{*}{Aorta} & Dice &  0.91 $\pm$ 0.04  & 0.91 $\pm$  0.03 & \textbf{\textcolor{blue}{0.92 $\pm$ 0.02}}\\
    \cline{2-2}
        & AHD & 0.77 $\pm$ 0.93  & 0.58 $\pm$ 0.67 & \textcolor{blue}{\textbf{0.52 $\pm$ 0.66}}\\
        \hline
            \multirow{2}{*}{Heart} & Dice &  0.93 $\pm$ 0.03 &0.93 $\pm$ 0.03 & 0.93 $\pm$ 0.03\\
    \cline{2-2}
        & AHD  & 0.25 $\pm$ 0.28 & 0.27 $\pm$  0.20 &  0.26 $\pm$ 0.22\\
        \hline
\end{tabular}
\end{table}

\subsection{Labeling issue}

\begin{figure}[t]
    \centering
    \includegraphics[width=0.8\textwidth]{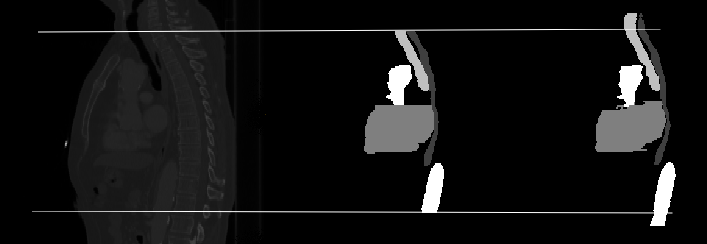}
    \caption{From left to right: CT scan of a patient, GT labeling and automatic segmentation with sU-Net in bilinear upsampling configuration. White: aorta, light gray: trachea, dark gray: heart, very dark gray: esophagus. Horizontal lines show lower and upper slice levels where GT stops. Note how the automated labeling is obtained beyond the limits and is thus counted as mislabeled, since GT is not present.}
    \label{fig:limits}
\end{figure}

Manual segmentation of the SegTHOR dataset is tailored according to the needs of radiotherapy and has not been performed for systematic segmentation evaluation. Thus, due to recommendations for manual segmentation, some slices located at the bottom or the top of the patient CT scan were not segmented. While this lack of manual labeling does not hinder the heart segmentation evaluation, this may be a problem for tubular organs which are perpendicular to the axial plane, such as the esophagus, the  trachea, and to a lesser extent, the aorta. For a majority of the 20 test patients, the automatic segmentation of the esophagus, trachea and aorta produced exceeds the upper and lower limits of manual segmentation as shown in Figure \ref{fig:limits}, and produces a labeling that is counted as missegmentation, since the corresponding ground truth (GT) does not exist.
We have thus run new experiments to assess the gain when evaluating on the restricted range of slices where the GT is present. From Table \ref{results_slices}, one can gather that for the esophagus, trachea and aorta, there is an improvement in Dice scores, especially for the trachea, but more significantly for the Average Hausdorff's distances. For future submission on the Codalab platform, we now offer two types of evaluation of the predicted segmentation: on all slices and on slices where the GT is present, i.e. by restricting the evaluation to a range of slices.\\

\begin{table}[t]
\caption{Average segmentation results ($\pm$ standard deviation) on the original dataset and on a dataset restricted to labeled slices only, with sU-Net with DR and bilinear upsampling. Best results are in bold. }
\label{results_slices}
\begin{center}

\begin{tabular}{|l|c||c|c|}
   \hline
    \multirow{2}{*}{OAR} &  & sU-Net & sU-Net  \\
     &  &original dataset & restricted dataset  \\
    \hline
    \multirow{2}{*}{Esophagus} & Dice &  0.81 $\pm$ 0.06 & \textbf{0.83 $\pm$ 0.06} \\
    \cline{2-2}
        & AHD &  0.68 $\pm$ 0.35& \textbf{0.32  $\pm$ 0.20} \\
    \hline
 \multirow{2}{*}{Trachea} & Dice & 0.86 $\pm$ 0.04 & \textbf{0.92 $\pm$ 0.02} \\
    \cline{2-2}
        & AHD &  1.08 $\pm$ 0.85 & \textbf{0.15 $\pm$ 0.09}  \\
        \hline
    \multirow{2}{*}{Aorta} & Dice & 0.92 $\pm$ 0.02 & \textbf{0.93 $\pm$ 0.02}\\
    \cline{2-2}
        & AHD &  0.52 $\pm$ 0.66& \textbf{0.19 $\pm$ 0.31}\\
        \hline
    \multirow{2}{*}{Heart} & Dice &   0.93 $\pm$ 0.03& 0.93 $\pm$ 0.03\\
    \cline{2-2}
        & AHD & 0.26 $\pm$ 0.22 &\textbf{0.16 $\pm$  0.15}\\
        \hline
   \end{tabular}

\end{center}
\end{table}

\section{Discussion and conclusion}

In this paper we have introduced SegTHOR, a dataset for the segmentation of organs at risk in CT images, available from the Codalab platform. The aim of the SegTHOR challenge is to foster research on this clinical application, but also to inspire the field of multilabel segmentation for (volumetric) anatomical images. We have presented several variants of a U-Net based architecture, that maybe used as first-line processing when dealing with a new medical image segmentation problem. Given the limited amount of data available, an architecture that is too deep and includes a large number of feature maps does not seem to be suitable for our semantic segmentation problem, in particular for the segmentation of the esophagus. We have presented a simplified CNN that was more appropriate to the problem at hand. Results show that the addition of the dropout has a major influence on the accuracy, and is a great help for most organs to improve the Dice metric as well as the AHD. In the decoding phase, the transposed convolution did not yield improved results compared to the bilinear upsampling operation; in this case, the bilinear interpolation should be favored to reduce computation time.

One limitation of our approach is that we only use one single reference segmentation. It is known that the variability of manual segmentation, be it intra- or inter-expert is not negligible. Most importantly, the OAR segmentation has a tremendous influence on dosimetric metrics \cite{vinod2016review}. Thus our next step will be to quantitatively assess the influence of OAR segmentation on dosimetric dose. In a study of a patient with oropharyngeal cancer \cite{nelms2012variations}, the authors found substantial dose differences resulting strictly from contouring variation, depending on the size, shape and location of the OAR. This emphasizes the need to accurately contour the OAR, in addition to the target tumor, when planning a radiotherapy. A dosimetric study would also allow to avoid the labeling issue present in the dataset.

Another use case of this dataset could be weakly supervised learning for image segmentation \cite{tajbakhsh2019embracing} or handling missing annotations \cite{petit2018handling}. Weakly supervised learning allows to reach full segmentation with partially annotated data, thus reducing the cost of full annotation. New challenges are arisen by this paradigm (how to leverage the weak labels? how to make use and model of external knowledge to help in the process?), which has been identified as a hot topic for the coming years \cite{tajbakhsh2019embracing}.






\section{Acknowledgements}
This project was co-financed by the European Union with the European regional development fund (ERDF, 18P03390/18E01750/18P02733) and by the Haute-Normandie R\'egional Council via the M2SINUM project. The authors would like to thank Prof. Carole Le Guyader (LMI, INSA Rouen) for her advice and the CRIANN (Centre des Ressources Informatiques et Applications Num\'erique de Normandie, France) for providing computational resources.

\section*{Disclosures}
No conflicts of interest, financial or otherwise, are declared by the authors.

\bibliographystyle{plain}
\bibliography{refs}

\end{document}